\documentclass[%
preprint%
 ,secnumarabic%
,superscriptaddress%
,amssymb, amsmath,nobibnotes,aps]{revtex4}

\usepackage{epsfig}%
\usepackage{graphicx}%
\expandafter\ifx\csname package@font\endcsname\relax\else
 \expandafter\expandafter
 \expandafter\usepackage
 \expandafter\expandafter
 \expandafter{\csname package@font\endcsname}%
\fi

\begin{document}

\title{Magnetic field generation in Higgs inflation model}
\author{Moumita Das}
\email{moumita@prl.res.in}
\author{Subhendra Mohanty}
\email{mohanty@prl.res.in}
\affiliation{Physical Research Laboratory, Ahmedabad 380009,
India}
\def\be{\begin{equation}}
\def\ee{\end{equation}}
\def\al{\alpha}
\def\bea{\begin{eqnarray}}
\def\eea{\end{eqnarray}}


\begin{abstract}
We study the generation of magnetic field in Higgs-inflation models where the Standard Model Higgs boson
has a large coupling to the Ricci scalar. We couple the Higgs field to the Electromagnetic fields via a non-
renormalizable dimension six operator suppressed by the Planck scale in the Jordan frame. We show that during Higgs inflation magnetic 
fields with present value $10^{-6}$ Gauss and comoving coherence length of $100 kpc$ can be generated in the Einstein frame. 
The problem of large back-reaction which is generic in the usual inflation models of magneto-genesis is avoided as the 
back-reaction is suppressed by the large Higgs-curvature coupling.

\end{abstract}
\maketitle

\section{Introduction} Observations \cite{Bernet, Kronberg} of magnetic field associated with high red-shift ($z>1$) galaxies suggest that
the large scale magnetic fields have a cosmological rather than an astrophysical origin. In the dynamo theory of magnetic field amplification in
galaxies \cite{Widrow, Brandenburg} and initial seed B- field of strength $10^{-20}$ G can be amplified to the observed $10^{-6}$ G by the
magnetohydrodynamics of galactic rotation. However observations \cite{Bernet, Kronberg} show that the magnetic fields associated with galaxies
have a very narrow spread around a micro-Gauss and therefore independent of the number of rotations of the galaxies. In the standard hot big bang
model of cosmology the generation of magnetic fields of large coherence scales ( 1kpc - 1 Mpc ) runs into problems with causality. For example if
B-fields are generated in the electro-weak era then the present coherence length of the 100 kpc would correspond to a length scale
$\lambda_{EW}=100 kpc (T_0/ 100 GeV)= 10^7 cm$. This length scale is much larger than the distance scale of causal Horizon at the electro-weak era
$H_{EW}^{-1} = 10^{-2} cm$. This suggests that if B-fields have an origin in the fundamental interaction then the perturbations must be super-horizon
which can happen during inflation. The magnetic field generated with the coherence scale $H_I$ during the time of inflation can be as large as the
present horizon $H_0$. Generation of magnetic field during inflation has been studied in extensively \cite{{Widrow},{Grasso},{Giovannini}} starting with
Turner and Widrow \cite{Turner} who coupled electromagnetic fields with curvature and the axion-inflaton and by Ratra \cite{Ratra} who coupled
electromagnetic field with the dilaton-inflaton. However in recent studies \cite{{Bamba},{Martin},{Demozzi},{Kanno}} it has been observed that in theories where
B-field is generated during the inflation the fluctuations of the electromagnetic field are as large as the perturbations of the inflaton
and spoil the  prediction of near-scale invariant primordial density perturbation of inflation.

A model of inflation with the standard model Higgs field has been proposed \cite{Bezrukov1,Bezrukov2} in which  the Higgs
has a large coupling with the Ricci scalar, $\xi \phi^2 R$ in the Jordan frame. This has the interesting
property that the Higgs potential $V=\lambda \phi^4$ in the Jordan frame transforms to a flat potential
$ \hat V \simeq \lambda M_P^4/(4 \xi^2)$  in the Einstein frame in the early universe (when $\phi > M_P/\xi$) which gives
almost scale invariant density perturbations $\Delta_R \sim 10^{-5}$ with the
parameters chosen as $\lambda =1$ and $\xi =4.6 \times 10^{4}$. In the
present epoch where $\phi=v=246$ GeV, the Ricci coupling term of the Higgs is negligibly small compared to the standard
 $M_P^2 R$ term of gravity and the Higgs potential in both Einstein and Jordan frames is $\lambda (\phi^2 -v^2)^2$ leading
 to the standard Higgs mass $m_h=\sqrt{2 \lambda} v$ with $\lambda \sim 1$ (instead of $\lambda \sim 10^{-13}$ as would be
 required to get the inflationary curvature perturbations of the right amplitude
 in an inflaton model with a $\lambda \phi^4$ potential without the added curvature coupling). The Higgs mass $\sqrt{2 \lambda}v$
 at scales $M_P/\xi$ must be renormalised down to the electroweak scale which leads to predictions for the Higgs mass
 to be in the range ($126$ Gev-$195$ Gev) \cite{Bezrukov3, Hertzberg}.

In this paper we introduce a non-renormalisable coupling of the Higgs with the electromagnetic fields of the form
$\frac{\phi^\dagger \phi}{M_P^2} F^2$. This is the leading order term in inverse powers of some large mass scale
(which we take to be $M_P$) which is invariant under the standard model symmetry group.   This term breaks the conformal 
symmetry and generates a magnetic field at the time of inflation when $\phi \sim M_P/\xi$. We find that the magnetic field 
generated at the time of inflation in terms of the Hubble parameter $\hat H$ is $\delta_B \sim {\hat H}^2 \xi$. This can be 
compared with fluctuation of the electromagnetic energy density $\rho_{EM}= (\phi^2/M_P^2) \, \delta_B^2 \sim \hat H^4$. 
The main point we illustrate in this paper is that if one starts with a conformal symmetry breaking electromagnetic interaction 
in the Jordan frame then one can generate a larger magnetic field by a factor of $\sqrt{\xi}$ in the Einstein frame and in addition 
the electromagnetic backreaction is smaller compared to the magnetic field. We show that with $\hat H\sim 10^{13}$ GeV and 
$\xi=4.6\times 10^{4}$ as required by the observations of the primordial density perturbations, one can generate magnetic fields 
of order $10^{-6}$ Gauss in the present universe with coherence length of $100$ kpc.

\section{Higgs-Inflation}
Consider the Higgs scalar $\Phi=\frac{1}{\sqrt{2}}(0,v+\phi)^T$ (in the unitary gauge) with a non-minimal coupling to the Ricci scalar, a Higgs potential $V(\phi)$ and a Higgs-photon interaction,
\begin{eqnarray}
S_{J}=\int d^{4}x\,\, \sqrt{-g}\left[-\frac{M_{p}^{2}+\xi \phi^{2}}{2}R+\frac{1}{2}g^{\mu\nu}\phi_{,\mu}\phi_{,\nu}
-V\left(\phi\right)-\frac{1}{4}I^{2}(\phi)\,F^{\mu\nu}F_{\mu\nu}\right]
\end{eqnarray}
where $V(\phi)= \frac{\lambda}{4}\left(\phi^{2}-v^{2}\right)$ Higgs potential with $v=246$ Gev and
 $I$ is the inverse of the electromagnetic coupling $I=1/g$.
 Here we assume
$I(\phi)$ has the explicit dependence on $\phi$ as,
\begin{eqnarray}
I^{2}(\phi)&=& \frac{\phi^{\dagger}\phi}{M_{p}^2}
\label{I_phi}
\end{eqnarray}
The metric in this `Jordan frame' (the frame in which there is a non-minimal Ricci coupling with the Higgs) is assumed to be,
\begin{eqnarray}
ds^{2}=dt^{2}-a^{2}(t) \delta_{ij}dx^{i}dx^{j} 
\end{eqnarray}

The non-minimal coupling with the gravity can be removed  by making a conformal transformation
to the `Einstein frame' \cite{{conf1},{conf2},{conf3}},
\begin{eqnarray}
g_{\mu\nu}\rightarrow \hat{g}_{\mu\nu}=\Omega^{2}g_{\mu\nu}\,\,\,\,\,\,  \rm{where} \,\,\,\,\Omega^{2}=1+\frac{\xi\phi^{2}}{M_{p}^{2}}
\label{con_trans}
\end{eqnarray}
In Einstein frame, the action will look like,
\begin{eqnarray}
S_{E}=\int d^{4}x\,\, \sqrt{-\hat{g}}\left[-\frac{M_{p}^{2}}{2}\hat{R}+\frac{1}{2\Omega^{2}}{g}^{\mu\nu}{\phi}_{,\mu}{\phi}_{,\nu}
+\frac{3\xi^{2}}{M_{p}^{2}\Omega^{4}}(\phi\,\phi_{,\mu})^{2}-\frac{V(\phi)}{\Omega^{4}}-\frac{1}{4}I^{2}(\phi)\,F^{\mu\nu}F_{\mu\nu}\right]
\label{SE1}
\end{eqnarray}
This conformal transformation produces non-canonical kinetic terms  for the scalar field $\phi$.
To make the kinetic term of the $\phi$ field canonical, we have to redefine the $\phi$ field in terms of new scalar,
\begin{eqnarray}
\frac{d\hat{\phi}}{d\phi}=\sqrt{\frac{\Omega^{2}+\frac{6\xi^{2} \phi^{2}}{M^{2}_{p}}}{\Omega^{4}}}
\label{new.phi}
\end{eqnarray}
The the action (\ref{SE1}), in terms of $\hat{\phi}$ is ,
\begin{eqnarray}
S_{E}=\int d^{4}x\,\, \sqrt{-\hat{g}}\left[-\frac{M_{p}^{2}}{2}\hat{R}+\frac{1}{2}\hat{g}^{\mu\nu}\hat{\phi}_{,\mu}\hat{\phi}_{,\nu}
-  \frac{V(\hat \phi)}{\Omega^{4}}   -\frac{1}{4}I^{2}(\hat{\phi})\,F^{\mu\nu}F_{\mu\nu}\right]
\end{eqnarray}
where $\phi$ is an implicit function of $\hat{\phi}$.

 When $\phi \ll M_P/\xi$ , $\Omega \simeq 1$ and $\hat{\phi}=\phi$. This corresponds to the situation in the present era where $\phi=v$. Inflation takes place when  $\phi \gg M_P/\xi$ and in this regime
$\Omega\simeq \sqrt{\xi}\phi / M_P$ and the relation between $\phi$ and $\hat{\phi}$  obtained from (\ref{new.phi}) is,
\begin{eqnarray}
\phi= \frac{M_{p}}{\sqrt{\xi}}\,\, \exp\left(\frac{\hat{\phi}}{\sqrt{6}M_{p}}\right)
\label{phi_phi_hat}
\end{eqnarray}

And the Higgs potential in this limit will be,
\begin{eqnarray}
{\hat V}=\frac{V}{\Omega^{4}}  \simeq \frac{\lambda \phi^{4}}{4\Omega^{4}} =\frac{\lambda M_{p}^{4}}{4 \xi^{2}}\left(1+\frac{M_{p}^{2}}{\xi \phi^{2}}\right)^{-2}
\label{v_prime}
\end{eqnarray}
Now using  equation~(\ref{phi_phi_hat}), we can write $ \hat{V}$ in terms of $\hat{\phi}$ as follows,
\begin{eqnarray}
\hat{V} =\frac{\lambda M_{p}^{4}}{4 \xi^{2}}\left(1+\exp \left(-\frac{2\hat{\phi}}{\sqrt{6}M_{p}}\right)\right)^{-2}
\end{eqnarray}
Since, $\phi > \frac{M_{p}}{\sqrt{\xi}}$, the Higgs-inflaton has the exponentially flat potential in the Einstein frame.

Due to the conformal transformation, the metric becomes,
\begin{eqnarray}
d\hat{s}^{2}=\Omega^{2}d{s}^{2}=d\hat{t}^{2}-\hat{a}^{2}(\hat{t}) \delta_{ij}dx^{i}dx^{j} 
\end{eqnarray}
where $\hat{a}=\Omega a$ and $d\hat{t}=\Omega dt$. 

The inflaton field satisfies the following equation,
\begin{eqnarray}
\frac{d\hat {\phi}}{d\hat{t}}&=&-\frac{\hat{V}'}{3\hat H}
\label{V_phi_eq1}
\end{eqnarray}
where prime denote the derivative with respect to $\hat{\phi}$. 
Solving this equation, we can find  the relation between $\phi$ and scale factor $\hat{a}$,
\begin{eqnarray}
\phi=\frac{2M_{p}}{\sqrt{3\xi}}(\log\hat{a})^{1/2}
\end{eqnarray}
Using the  equation~(\ref{I_phi}) 
, we can write $I(\phi)$ as a function of $`\hat{a}$' as follows,
\begin{eqnarray}
I(\phi)=\frac{2}{\sqrt{3\xi}}(\log\hat{a})^{1/2}
\label{rel.I}
\end{eqnarray}

For the calculation of $\delta_{B}$, we have to know the value of $\xi$, which can be calculated from the curvature
perturbation. The amplitude of the curvature perturbation can be written as,
\begin{eqnarray}
\Delta^{2}_{R}=\frac{1}{4\pi^{2}}\left(\frac{\hat{H}^{2}}{d\hat{\phi}/d\hat{t}}\right)^{2}=\frac{3}{8\pi^{2}}\frac{\hat{H}^{4}}{\epsilon}
\label{curve_perturb}
\end{eqnarray}
where, $H^{2}=\frac{8\pi G}{3}\hat{V}=\frac{\hat{V}}{3M^{2}_{P}}$, $d\hat{\phi}/d\hat{t}=-\hat{V}'/3\hat{H}$ and
$\epsilon=\frac{M^{2}_{p}}{2}\left(\frac{\hat{V}'}{\hat{V}}\right)^{2}\simeq\frac{4M^{4}_{p}}{3\xi^{2}\phi^{4}}$.
Therefore, the amplitude of the curvature perturbation becomes,
\begin{eqnarray}
\Delta^{2}_{R}=\frac{1}{24\pi^{2}}\frac{\hat{V}}{\epsilon}\frac{1}{M^{4}_{p}}=5.23\frac{\lambda}{\xi^{2}}
\end{eqnarray}
where we have taken $\hat{V}\simeq\frac{\lambda M^{4}_{p}}{4\xi^{2}}$ and $\phi=9.01\frac{M_{p}}{\sqrt{\xi}}$ for $N=61$.
Taking $\lambda=1$ and the WMAP result \cite{Jarosik}
 $\Delta^{2}_{R}=2.43\times10^{-9}$,  we
see that  $\xi=4.6 \times 10^4$ . One  can check  the value of $\xi$ which gives the correct amplitude of curvature
perturbation predicts that the  spectral index $n_{s}$ is,
\begin{eqnarray}
n_{s}=1+2\eta-6\epsilon=0.965
\end{eqnarray}
(where $\eta=M^{2}_{p}\left(\frac{\hat{V}''}{\hat{V}}\right)\simeq-\frac{4M^{2}_{p}}{3\xi\phi^{2}}$) which is consistent 
with the WMAP result $ n_s=0.963 \pm 0.012$ \cite{Jarosik}.
\section{Generation of Magnetic field during Higgs-Inflation }
We consider the term containing massless vector field separately from the Einstein action,
\begin{eqnarray}
S_{E}=\int d^{4}x\,\, \sqrt{-\hat{g}}\left[-\frac{1}{4}I^{2}(\hat \phi)\,F^{\mu\nu}F_{\mu\nu}\right]
\end{eqnarray}
where $F_{\mu\nu}=\partial_{\mu}A_{\nu}-\partial_{\nu}A_{\mu}$ and $A_{\mu}=\left(A_{0},A_{i}\right)$.
Decomposing the spatial part $A_{i}$ in terms of its transverse and longitudinal components
$A_{i}=A^{T}_{i}+\partial_{i}\chi$ and considering $\partial_{i}A^{T}_{i}=0$ and $A_{0}=\chi\prime$, we get
the action as follows,
\begin{eqnarray}
S_{E}=\int d^{4}x\,\,I^{2}\left(A^{T\prime}_{i} A^{T\prime}_{i}+A^{T}_{i}\Delta A^{T}_{i}\right)
\end{eqnarray}
where primes denotes the derivative with respect to the conformal time $\tau$.
The transverse component of $A_{i}$ can written in Fourier space as,
\begin{eqnarray}
A^{T}_{i}(x,\hat{\tau})=\sum_{\sigma=1,2} \int \frac{d^{3}k}{(2\pi)^{3/2}}\,\,A^{\sigma}_{\textbf{k}}(\hat{\tau})\,
\varepsilon^{\sigma}_{i}(\textbf{k}) e^{i\textbf{k}\cdot x}
\end{eqnarray}
where $\varepsilon^{\sigma}_{i}(\textbf{k}),\sigma=1,2$ are two orthogonal polarization vectors and satisfying the relations
$k_{i}\varepsilon^{\sigma}_{i}(\textbf{k})=0$ and $\varepsilon^{\sigma}_{i}(-\textbf{k})\varepsilon^{\rho}_{i}
(\textbf{k})=\delta^{\sigma\rho}$, the action will be,
\begin{eqnarray}
S_{E}= \frac{1}{2}\sum_{\sigma=1,2} \int I^{2}\left(A^{\sigma\prime}_{\textbf{k}}A^{\sigma\prime}_{-\textbf{k}}-k^{2}A^{\sigma}_{\textbf{k}}A^{\sigma}_{-\textbf{k}}\right)d\hat{\tau} d^{3}k
\end{eqnarray}
Defining $A^{\sigma}_{\textbf{k}}=\frac{\tilde{A}^{\sigma}_{\textbf{k}}}{I}
$, the action becomes,
\begin{eqnarray}
S_{E}= \frac{1}{2}\sum_{\sigma=1,2} \int\left[\tilde{A}^{\sigma\prime}_{\textbf{k}}\tilde{A}^{\sigma\prime}_{-\textbf{k}}-\left(k^{2}-\frac{I''}{I}\right)\tilde{A}^{\sigma}_{\textbf{k}}\tilde{A}^{\sigma}_{-\textbf{k}}\right]d\hat{\tau} d^{3}k
\end{eqnarray}
We can expand $\tilde{A}^{\sigma}_{\textbf{k}}$ in terms of the creation and annihilation operators as follows,
\begin{eqnarray}
\tilde{A}^{\sigma}_{\textbf{k}}= \frac{1}{\sqrt2}\left(u_{\textbf{k}}a^{\sigma}_{\textbf{k}}+u^{\star}_{\textbf{k}}a^{\sigma\dagger}_{\textbf{k}}\right)
\label{rel.v.u}
\end{eqnarray}
where the creation and annihilation operators satisfies $\left[a^{\sigma}_{\textbf{k}},a^{\rho\dagger}_{\textbf{k}^{\prime}}\right]
=\delta^{\sigma\rho}\delta(\textbf{k}-\textbf{k}^{\prime})$.

Therefore, $u_{\textbf{k}}$ will satisfy the equation,
\begin{eqnarray}
u^{\prime\prime}_{\textbf{k}}+\left(k^{2}-\frac{I^{\prime\prime}}{I}\right)u_{\textbf{k}}=0
\label{e.o.m}
\end{eqnarray}
For large value of $k$, equation~(\ref{e.o.m}) reduces to
\begin{eqnarray}
u^{\prime\prime}_{\textbf{k}}+k^{2}u_{\textbf{k}}=0
\end{eqnarray}
and the solution will be of the form,
\begin{eqnarray}
u_{\textbf{k}\,_{>}}=\frac{1}{\sqrt{2k}}e^{i k\hat{\tau}}
\label{large.k}
\end{eqnarray}
But for smaller value of $k$ , the term $\frac{I''}{I}$ will dominate and the solution will be,
\begin{eqnarray}
u_{\textbf{k}\,_{<}}=c_{1}I+c_{2}I\int\frac{d\hat{\tau}}{I^{2}}
\label{1.small.k}
\end{eqnarray}
Using the relation $\hat{\tau}=-1/\hat{a}\hat{H}$ and the expression of $I$ in equation~(\ref{rel.I}), equation~(\ref{1.small.k}) can be simplified to
\begin{eqnarray}
u_{\textbf{k}\,_{<}}&=&\frac{2c_{1}}{\sqrt{3\xi}}(\log\hat{a})^{1/2}-\frac{c_{2}}{2\hat{H}}\frac{\sqrt{3\xi}}{\hat{a}(\log\hat{a})^{1/2}}\nonumber\\
&\simeq&\frac{2c_{1}}{\sqrt{3\xi}}(\log\hat{a})^{1/2}
\label{2.small.k}
\end{eqnarray}
where second term is neglected as it is suppressed  by the factor of $\hat a\, (\log\hat{a})^{1/2}$ in the denominator. 
By matching the equation~(\ref{large.k}) and equation~(\ref{2.small.k}) 
at $\hat{\tau}=\frac{1}{k}$, we determine the constant $c_{1}$,
\begin{eqnarray}
c_{1}&=&\frac{\sqrt{3}e^{i}}{2}\sqrt{\frac{\xi}{2k\log\hat{a}_{k}}}\nonumber
\end{eqnarray}
where $\hat{a}_{k}=\frac{k}{\hat{H}}$ is the scale factor at $\hat{\tau}=\frac{1}{k}$.
Therefore the solution of the mode functions of the electromagnetic perturbations are of the form,
\begin{equation}
u_{\textbf{k}}\simeq\frac{e^{i}}{\sqrt{2k}}\sqrt{\frac{\log\hat{a}}{\log\hat{a}_{k}}}
\label{final.v}
\end{equation}
The correlation function will be,
\begin{eqnarray}
<0|\hat{{A}}^{T}_{i}(\hat{\tau},x) \hat{{A}}^{Ti}(\hat{\tau},y)|0>&=&\frac{1}{a^{2}I^{2}}\sum_{\sigma\sigma^{\prime}}\int\frac{d^{3}k\,d^{3}k^{\prime}}{\left(2\pi\right)^{3}}e^{i(\textbf{k}\cdot\textbf{x}+\textbf{k}\cdot\textbf{y})}<0|u^{\sigma}_{\textbf{k}}u^{\sigma\prime}_{\textbf{k}\prime}|0>\nonumber\\
&=&\frac{1}{4\pi^{2}\hat{a}^{2}I^{2}}\int\frac{dk}{k}\,\,|u_{\textbf{k}}|^{2} k^{3}\,\,\frac{\sin{k(x-y)}}{k(x-y)} \nonumber\\
&\equiv&\int\frac{dk}{k}\,\,\delta^{2}_{A}(k,\hat{\tau})\,\,\frac{\sin{k(x-y)}}{k(x-y)}
\end{eqnarray}
The power spectrum of the vector field $\delta^{2}_{A}(k,\hat{\tau})$ can be identified with,
\begin{eqnarray}
\delta^{2}_{A}(k,\hat{\tau})=\frac{|u_{\textbf{k}}|^{2} k^{3}}{4\pi^{2}\hat{a}^{2}I^{2}}
\label{A.spectra}
\end{eqnarray}
Using the relation between magnetic field and vector field $B^{2}=\frac{1}{2a^{4}}F_{ik}F_{ik}=\frac{1}{a^{4}}\left(\partial_{i}A_{k}\partial_{i}A_{k}-\partial_{k}A_{i}\partial_{k}A_{i}\right)$,
we can calculate the power spectrum of the magnetic field $\delta^{2}_{B}(k,\hat{\tau})$ as follows,
\begin{eqnarray}
\delta^{2}_{B}(k,\hat{\tau})=\delta^{2}_{A}(k,\hat{\tau})\frac{k^{2}}{\hat{a}^{2}}=\frac{|u_{\textbf{k}}|^{2} k^{5}}{4\pi^{2}\hat{a}^{4}I^{2}}
\label{B.spectra}
\end{eqnarray}
Using the expression for $u_{\textbf{k}}$ from equation(\ref{final.v}), we can calculate $\delta_{B}$, at the time of horizon
crossing as follows,
\begin{equation}
\delta^{2}_{B}(k)\simeq\frac{3}{32\pi^{2}}\frac{\hat{H}^{4}\xi}{|\log {\frac{k}{\hat H}}|}
\end{equation}
 At the time of inflation $\delta^{2}_{B}=1.9\times10^{53}\,GeV^{4}$ for modes of with a co-moving coherence length $k^{-1}=100$ kpc.  
After horizon exit $\delta^{2}_{B}$ varies as $\frac{1}{\hat{a}^{4}}$,so we can calculate $\delta^{2}_{B_0}$ at present. 
If $N$ is the no of e-foldings after the some specific mode (like $k=100$ kpc), leaves the de-Sitter horizon, 
then $\delta^{2}_{B_0}$ and $\delta^{2}_{B}$ are related as follows,
\begin{eqnarray}
\delta^{2}_{B_0}=\delta^{2}_{B}\left(\frac{\hat{a}_{I}}{\hat{a}_{0}}\right)^{4}=\delta^{2}_{B_I}\exp(-4N)
\end{eqnarray}
And we find that $N=61$ gives the value of magnetic perturbation at the present epoch,
$\delta_{B_0}=1.5\times10^{-26}\,GeV^{2}= 1.7 \times 10^{-6} $ Gauss at length scales of $100 $kpc.
We have to study the back reaction of the generated electromagnetic field on the background. For this, we will calculate the
energy density $\rho_{em}$ which is defined as $T^{0}_{0}$ component of the energy-momentum tensor.
\begin{eqnarray}
T^{0}_{0}=I^{2}\left(\frac{1}{4}F_{\alpha\beta}F^{\alpha\beta}-F_{0\alpha}F^{0\alpha}\right)=\frac{I^{2}}{2\hat{a}^{4}}\left(A^{T\prime}_{i}A^{T\prime}_{i}+\partial_{i}A^{T}_{k}\partial_{i}A^{T}_{k}\right)
\end{eqnarray}
Using the relation 
(\ref{rel.v.u}) , the energy density $\rho_{em}$ will be,
\begin{eqnarray}
\rho_{em}=<0|\hat{T}^{0}_{0}|0>&=&\frac{1}{8\pi^{2}\hat{a}^{4}}\int \frac{dk}{k}\,\,\left(|u^{\prime}_{\textbf{k}}|^{2}+k^{2}|u_{\textbf{k}}|^{2}\right)k^{3}
\label{rho}
\end{eqnarray}
neglecting the derivatives of $I$.
Using the solution (\ref{final.v}) for the mode functions expression in equation(\ref{rho}), the energy density perturbations of electromagnetic fields,
\begin{eqnarray}
\rho_{em}&=&\frac{\hat{H}^{4}}{64\pi^{2}}+\frac{\hat{H}^{4}}{8\pi^{2}}\left[\frac{1}{16\log{\hat{a}}}\left(\frac{1}{(\log{\hat{a}})^2}+\frac{2}{\log{\hat{a}}}-\left(\frac{\hat{a}_{i}}{\hat{a}}\right)^{2}\frac{2}{\log{\hat{a}_{i}}}-\left(\frac{\hat{a}_{i}}{\hat{a}}\right)^{2}\frac{1}{(\log{\hat{a}_{i}})^2}\right)\right.\nonumber\\
&+&\left.\frac{\log{\hat{a}}}{32}\left(\frac{1}{(\log{\hat{a}})^2}-\left(\frac{\hat{a}_{i}}{\hat{a}}\right)^{4}\frac{4}{\log{\hat{a}_{i}}}-\left(\frac{\hat{a}_{i}}{\hat{a}}\right)^{4}\frac{1}{(\log{\hat{a}_{i}})^2}\right)\right]\nonumber\\
&\simeq&8.01\times10^{49}\,\,GeV^{4}
\label{ED_super}
\end{eqnarray}
We see that the energy density of the electromagnetic fields (\ref{ED_super}) is smaller than the perturbation of the magnetic field $\delta_B^2 \simeq 10^{53}$ Gev$^4$. This is due to the fact that the electromagnetic coupling 
$g=1/I$ becomes large (of the order 10) in the Einstein frame during inflation. This means that  perturbative calculations in expansion of $g$ are invalid in the regime $\phi> M_P/\xi$ which holds during inflation.


\section{Conclusion}
In this paper we have shown that the Higgs model of inflation \cite{Bezrukov1,Bezrukov2} in which a large Higgs-Ricci coupling gives rise to a 
flat Higgs potential in the Einstein frame in early universe, is also ideal for generation of magnetic field during inflation. Breaking the 
conformal invariance of electromagnetism by a non-renormalizable Higgs-photon coupling term in the Jordan frame enables us to generate large scale 
magnetic field during inflation
while keeping the backreaction pointed out in \cite{{Bamba},{Martin},{Demozzi}, {Kanno}} under control.
The consequences of primordial magnetic field fluctuations on the CMB anisotropy has been studied in \cite{Yamazaki,Dulaney}. The cosmological 
isotropy is broken by large scale magnetic fields which will show up in the CMB anisotropy and polarization spectrum. This points to the possibility 
that the magnetic field generation model studied in this paper can be tested in the forthcoming CMB anisotropy measurement experiments 
like PLANCK \cite{Planck}.



\begin{thebibliography}{}
\bibitem{Bernet}
  M.~L.~Bernet, F.~Miniati, S.~J.~Lilly, P.~P.~Kronberg and M.~Dessauges-Zavadsky,
  Nature {\bf 454}, 302 (2008)
  [arXiv:0807.3347 [astro-ph]].

 \bibitem{Kronberg}
  P.~P.~Kronberg, M.~L.~Bernet, F.~Miniati, S.~J.~Lilly, M.~B.~Short and D.~M.~Higdon,
  Astrophys.\ J.\  {\bf 676}, 7079 (2008)
  [arXiv:0712.0435 [astro-ph]].

\bibitem{Brandenburg}
  A.~Brandenburg and K.~Subramanian,
  Phys.\ Rept.\  {\bf 417}, 1 (2005)
  [arXiv:astro-ph/0405052].



\bibitem{Widrow}
  L.~M.~Widrow,
  Rev.\ Mod.\ Phys.\  {\bf 74}, 775 (2002)
  [arXiv:astro-ph/0207240].




\bibitem{Grasso}
  D.~Grasso and H.~R.~Rubinstein,
  Phys.\ Rept.\  {\bf 348}, 163 (2001)
  [arXiv:astro-ph/0009061].


\bibitem{Giovannini}
  M.~Giovannini,
  Int.\ J.\ Mod.\ Phys.\  D {\bf 13}, 391 (2004)
  [arXiv:astro-ph/0312614].



\bibitem{Turner}
  M.~S.~Turner and L.~M.~Widrow,
  Phys.\ Rev.\  D {\bf 37}, 2743 (1988).

\bibitem{Ratra}
  B.~Ratra,
  Astrophys.\ J.\  {\bf 391}, L1 (1992).
  
\bibitem{Bamba}
  K.~Bamba, N.~Ohta and S.~Tsujikawa,
  Phys.\ Rev.\  D {\bf 78}, 043524 (2008)
  [arXiv:0805.3862 [astro-ph]].
  


\bibitem{Martin}
  J.~Martin and J.~Yokoyama,
  JCAP {\bf 0801}, 025 (2008)
  [arXiv:0711.4307 [astro-ph]].


\bibitem{Demozzi}
  V.~Demozzi, V.~Mukhanov and H.~Rubinstein,
  JCAP {\bf 0908}, 025 (2009)
  [arXiv:0907.1030 [astro-ph.CO]].

\bibitem{Kanno}
  S.~Kanno, J.~Soda and M.~a.~Watanabe,
  JCAP {\bf 0912}, 009 (2009)
  [arXiv:0908.3509 [astro-ph.CO]].



\bibitem{Bezrukov1}
  F.~L.~Bezrukov and M.~Shaposhnikov,
  Phys.\ Lett.\  B {\bf 659}, 703 (2008)
  [arXiv:0710.3755 [hep-th]].

\bibitem{Bezrukov2}
  F.~Bezrukov, D.~Gorbunov and M.~Shaposhnikov,
  JCAP {\bf 0906}, 029 (2009)
  [arXiv:0812.3622 [hep-ph]].



\bibitem{Bezrukov3}
  F.~Bezrukov and M.~Shaposhnikov,
  JHEP {\bf 0907}, 089 (2009)
  [arXiv:0904.1537 [hep-ph]].

\bibitem{Hertzberg}
  A.~De Simone, M.~P.~Hertzberg and F.~Wilczek,
  Phys.\ Lett.\  B {\bf 678}, 1 (2009)
  [arXiv:0812.4946 [hep-ph]].


\bibitem{conf1}
  D.~I.~Kaiser,
  Phys.\ Rev.\  D {\bf 52}, 4295 (1995)
  [arXiv:astro-ph/9408044].

\bibitem{conf2}
  E.~Komatsu and T.~Futamase,
  Phys.\ Rev.\  D {\bf 58}, 023004 (1998)
  [arXiv:astro-ph/9711340].
\bibitem{conf3}
  E.~Komatsu and T.~Futamase,
  Phys.\ Rev.\  D {\bf 59}, 064029 (1999)
  [arXiv:astro-ph/9901127].

\bibitem{Jarosik}
  N.~Jarosik {\it et al.},
  arXiv:1001.4744 [astro-ph.CO].

\bibitem{Yamazaki}
  D.~G.~Yamazaki, K.~Ichiki, T.~Kajino and G.~J.~Mathews,
  Phys.\ Rev.\  D {\bf 77}, 043005 (2008)
  [arXiv:0801.2572 [astro-ph]].

\bibitem{Dulaney}
  T.~R.~Dulaney and M.~I.~Gresham,
  arXiv:1001.2301 [astro-ph.CO].

\bibitem{Planck}
    [Planck Collaboration],
  arXiv:astro-ph/0604069.



\end{thebibliography}
\end{document}